# Transition from phase slips to the Josephson effect in a superfluid $^4$He weak link

E. Hoskinson[1], Y. Sato[1], I. Hahn[2] and R. E. Packard[1]

[1]*Department of Physics, University of California, Berkeley, CA, 94720, USA.* [2]*Jet Propulsion Laboratory, California Institute of Technology, Pasadena, CA 91109, USA*

**The rich dynamics of flow between two weakly coupled macroscopic quantum reservoirs has led to a range of important technologies. Practical development has so far been limited to superconducting systems, for which the basic building block is the so-called superconducting Josephson weak link[1]. With the recent observation of quantum oscillations[2] in superfluid $^4$He near 2K, we can now envision analogous practical superfluid helium devices. The characteristic function which determines the dynamics of such systems is the current-phase relation $I_s(\varphi)$, which gives the relationship between the superfluid current $I_s$ flowing through a weak link and the quantum phase difference $\varphi$ across it. Here we report the measurement of the current-phase relation of a superfluid $^4$He weak link formed by an array of nano-apertures separating two reservoirs of superfluid $^4$He. As we vary the coupling strength between the two reservoirs, we observe a transition from a strongly coupled regime in which $I_s(\varphi)$ is linear and flow is limited by $2\pi$ phase slips, to a weak coupling regime where $I_s(\varphi)$ becomes the sinusoidal signature of a Josephson weak link.**

The dynamics of flow between two weakly coupled macroscopic quantum reservoirs can be highly counterintuitive. In both superconductors and superfluids, currents will oscillate through a constriction (weak link) between two reservoirs in response to a static driving force which, in a classical system, would simply yield flow in one direction. In superconductors, such junctions have given rise to a range of technologies. Although promising analogous devices[3,4,5] based on weak links have been demonstrated in superfluid $^3$He, practical development will be hampered by the difficulty of working at the very low temperatures ($<10^{-3}$K) required. Quantum oscillations were recently observed in superfluid $^4$He at a temperature 2000 times higher. To understand the fundamental nature of these oscillations, and to make progress toward device development, it is necessary to know the relationship between current and phase difference across the junction, $I_s(\varphi)$. The measurement of $I_s(\varphi)$ reported here reveals a transition between two distinct quantum regimes, and opens the door to the development of superfluid $^4$He interference devices analogous to the dc-SQUID, which will be highly sensitive to rotation.

When superfluid $^4$He, well below its transition temperature $T_\lambda$ = 2.17 K, is forced through a constriction, it will accelerate until it reaches a critical velocity, $v_c$, at which a quantized vortex is nucleated. This is shown schematically in Fig. 1. The vortex moves across the path of the fluid, decreasing the quantum phase difference between the reservoirs by $2\pi$ and decreasing the fluid velocity[6] by a quantized amount $\Delta v_s$. This phase slip process repeats, such that the flow through the constriction follows a sawtooth waveform. The critical velocity decreases toward zero as T is increased towards $T_\lambda$, but $\Delta v_s$ is mostly independent of T. When $v_c < \Delta v_s$, the flow actually reverses direction whenever a phase slip occurs. If this situation were to continue as $T \rightarrow T_\lambda$ and $v_c$ drops below $\Delta v_s/2$, upon phase slipping the flow would end up with a velocity greater than $v_c$ in the opposite direction. This could not be an energy conserving process. At about the same temperature that this would occur, the healing length of the superfluid, $\xi_4/d = 0.34(1-T/T_\lambda)^{-0.67}$ nm, becomes comparable to the diameter of the constriction, $d$. Superfluidity is then suppressed in the confined geometry of the constriction, which now acts as a barrier between the two reservoirs of superfluid, analogous to a Dayem bridge in superconductors[7]. In this limit, where the wave functions on either side of the barrier partially overlap, the dynamics of flow through the aperture is expected to be described by the Josephson effect equations, which predict sinusoidal, rather than sawtooth, oscillations. Previously, flow features of a hydrodynamic resonator were found to be consistent with such sinusoidal behavior[8]. The system can be brought from one limit to the other by varying the strength of the coupling through the aperture. This coupling strength depends on the ratio $\xi_4/d$ and we control $\xi_4$ by varying T.

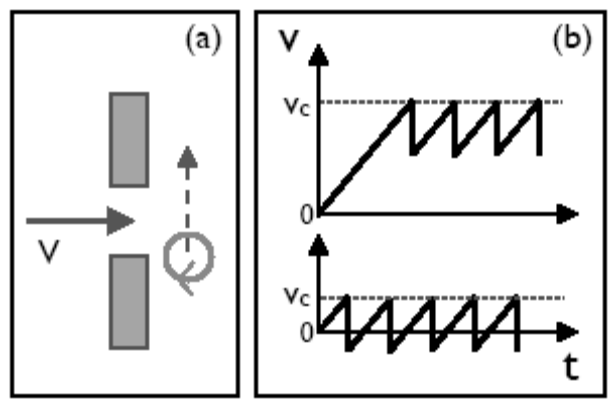

**Figure 1.** Schematic of flow through an aperture and corresponding velocity evolution. Superfluid with velocity v accelerates in response to a driving force up to a critical velocity $v_c$ at which a singly quantized vortex is nucleated, crosses the flow path, and causes a drop in v. Repeated vortex nucleation events give rise to a sawtooth waveform. The critical velocity $v_c$ drops as $T \rightarrow T_\lambda$. At some T (lower curve, part b) the superfluid flow will actually reverse direction.

A schematic of our experimental cell, described in more detail elsewhere[9], is shown in Fig 2a. A cylindrical inner reservoir of diameter 8mm and height 0.6mm is bounded on the top by an 8μm thick flexible Kapton diaphragm on which a 400nm thick layer of superconducting lead has been evaporated. An array of 4225 apertures spaced on a 3μm square lattice in a 50nm thick silicon nitride membrane is mounted in a rigid aluminum plate forming the walls and bottom of the inner reservoir. Flow measurements both above and below $T_\lambda$ indicate that the apertures are $d = 38 \pm 9$ nm in diameter. Pressures can be induced across the array by application of an electrostatic force between the diaphragm and a nearby electrode, thereby pulling up on the diaphragm. Above the electrode is a superconducting coil (not shown) in which a persistent electrical current flows, producing a magnetic field which is modified by the superconducting plane of the diaphragm. Motion of the diaphragm, indicating fluid flow through the array, induces changes in the persistent current flowing in the coil, which are detected with a SQUID. The output of the SQUID is proportional to the displacement of the diaphragm $x(t)$. We can resolve a displacement of $2 \times 10^{-15}$ m in one second.

The inner reservoir sits in a sealed can filled from room temperature with $^4$He through a capillary which, in order to decouple the can and inner reservoir from environmental fluctuations, is then blocked close to the can with a cryogenic valve. The can is immersed in a pumped bath of liquid helium which is temperature stabilized to within $\pm 50$ nK using a distributed resistive heater and high resolution thermometer[10] in a feedback loop.

  

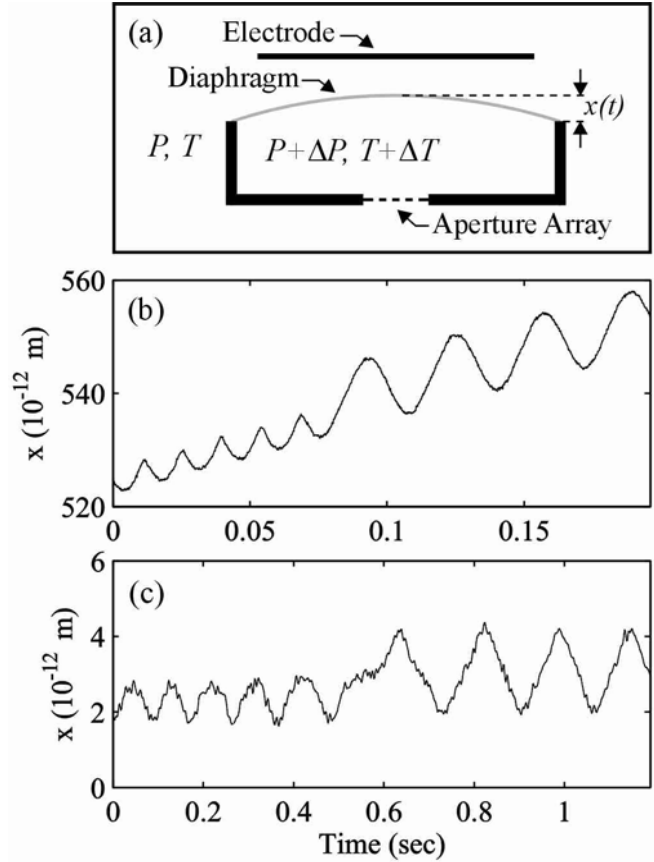

**Figure 2**. Schematic of the experimental cell and two typical flow transients. (a) Schematic. (b), (c) Diaphragm position $x(t)$ as a function of time within two flow transients. These plots each show both lower amplitude Josephson frequency oscillations, for which $\langle \Delta\mu \rangle > 0$, and larger amplitude Helmholtz oscillations, for which $\langle \Delta\mu \rangle = 0$. In (b), $T_\lambda - T = 7.4\,\text{mK}$, $\xi_4/d = 0.4$ and $I_s(\varphi)$ is mostly linear. The Josephson frequency oscillations occur by the phase slip mechanism here – the superfluid velocity, proportional to $dx/dt$, follows a sawtooth waveform. In (c), $T_\lambda - T = 0.8\,\text{mK}$, $\xi_4/d = 1.8$ and $I_s(\varphi)$ is mostly sinusoidal. Near $t = 0.6$ sec where the Josephson frequency oscillations decay into Helmholtz oscillations, $\varphi$ approaches a local maximum of just less than π. The current $I \propto dx/dt$ slows here, reflecting the sinusoidal nature of $I_s(\varphi)$.

Superfluid $^4$He with superfluid density $\rho_s$ is described by a macroscopic quantum wave function $\psi = \sqrt{\rho_s}\,e^{i\phi}$. In unrestricted space, the flow velocity is proportional to the gradient of the phase: $v_s = (\hbar/m_4)\nabla\phi$. Here $\hbar$ is Plank's constant $h$ divided by $2\pi$ and $m_4$ is the $^4$He atomic mass. The superfluid current $I_s$ through our array is a function of the phase difference $\Delta\phi$ between the two reservoirs. In general this phase difference evolves according to the Josephson-Anderson phase evolution equation, $d\Delta\phi/dt = -\Delta\mu/\hbar$, where $\Delta\mu = m_4(\Delta P/\rho - s\Delta T)$ is the chemical potential difference across the array. Here $\rho$ is the fluid total mass density, $s$ is the entropy per unit mass, and $\Delta P$ and $\Delta T$ are the pressure and temperature differences across the array. If $I_s(\Delta\phi)$ is $2\pi$ periodic, a constant $\Delta\mu$ gives rise to oscillations at the Josephson frequency, $f_J = \Delta\mu/h$. This can occur in either the strong coupling phase slip regime or the weak coupling Josephson regime. Our goals here are to determine the detailed time evolution of these oscillations and the underlying current-phase relation $I_s(\Delta\phi)$ as we change the coupling strength by varying the temperature. Hereafter we use $\varphi$ to denote $\Delta\phi$.

In Fig 2, panels b and c are sections of two flow transients excited by a step in the pressure $\Delta P$ across the array. The curves show the displacement $x(t)$ of the diaphragm as fluid is driven though the apertures under the influence of a time dependent chemical potential gradient. The slope of these curves, $dx/dt$, is proportional to the total mass current through the aperture array, $I(t) = \rho A\, dx/dt$, where $A$ is the diaphragm area. For both flow transients, the pressure step is such that a conventional fluid would be driven in the positive direction (positive slope). In panel b, where $T_\lambda$-T = 7.4 mK, $\xi_4/d = 0.4$ and the coupling is relatively strong. The clear slope discontinuities are the signatures of phase slips that occur in the first half of the plot whenever the continuously accelerating flow reaches a maximum current $I_c$. By contrast in panel c, where $T_\lambda$-T = 0.8 mK, $\xi_4/d = 1.8$ and the system is in the Josephson weak link regime. The sharp phase slip discontinuities have been smoothed out into sinusoidal Josephson oscillations. The $\Delta\mu$ induced by the initial pressure step relaxes to equilibrium throughout each transient. When $\Delta\mu$ reaches zero the Josephson frequency oscillations cease and lower frequency resonant "Helmholtz" or "pendulum-mode" oscillations begin, with current amplitude less than $I_c$. These are the larger displacement oscillations in the second halves of panels b and c.

The method we use to determine $I_s(\varphi)$ is conceptually similar to that used[11] by Marchenkov et al for $^3$He. The superfluid current as a function of time $I_s(t)$ is determined from transient data such as that shown in Fig. 2, with a small correction due to a small normal flow component $I_n(t)$. The phase difference across the aperture array is determined by integrating the phase evolution equation: $\varphi(t) = \varphi(0) - \hbar^{-1} \int_0^t \Delta\mu(\tau) d\tau$, where the phase offset is determined by the fact that $\varphi = 0$ when $I_s = 0$. Elimination between $I_s(t)$ and $\varphi(t)$ of the common variable of time then yields the current-phase relation $I_s(\varphi)$.

Integration of the phase evolution equation requires knowledge of both $\Delta P(t)$ and $\Delta T(t)$. $\Delta P(t)$ is directly determined by the diaphragm displacement: $\Delta P = kx/A$, where $k$ is a measured spring constant. An absolute calibration of $\Delta P$ is provided by the Josephson frequency relation for $\Delta T = 0$: $\Delta P = \rho h f_J / m_4$ (with $k$ and $A$ this in turn provides the calibration for $x$). A temperature difference $\Delta T(t)$ is created whenever superfluid flows into or out of the inner cell (the thermo-mechanical effect) and is calculated using the measured current and a simple heat flow equation[9].

The current-phase functions for several temperatures are shown in Fig. 3. A smooth transformation occurs from the low temperature strong coupling regime where $I_s(\varphi)$ is linear with limiting values, into the weak coupling regime, within a few mK of $T_\lambda$, where $I_s(\varphi)$ morphs into a sine function. For $T_\lambda - T > 5$ mK, $I_s(\varphi)$ is mostly linear and the system is in the phase slip regime. Under the influence of a constant $\Delta\mu$, $\varphi$ will increase linearly until it reaches a critical value $\varphi_c$ (the maximum value of $\varphi$ for each plot) then slips back discontinuously by $2\pi$. $I_s$ drops from $I_s(\varphi_c)$ to $I_s(\varphi_c - 2\pi)$. Because $\varphi_c$ is less than $2\pi$ at these temperatures, $\varphi$ and $I_s$ reverse direction when a phase slip occurs. As the temperature is increased, going from top to bottom in Fig. 3, $\varphi_c$ decreases. Around the temperature at which $\varphi_c$ reaches $\pi$, $I_s(\varphi)$ morphs into a sinusoid, which is the signature of an ideal Josephson weak link. Each of the curves in

Fig. 3 is obtained by averaging the $I_s(\varphi)$ data from between 5 and 70 transients such as those in Fig. 2.

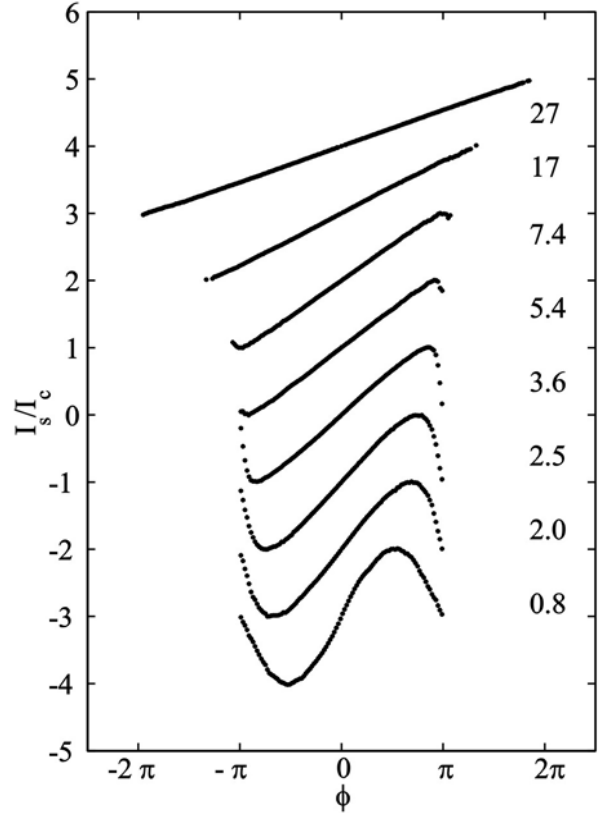

**Figure 3**. Evolution of $I_s(\varphi)$ with temperature near $T_\lambda$. Each curve has been normalized by its maximum value $I_c$ and shifted vertically. The corresponding $T_\lambda$-$T$ is indicated in mK to the right of each curve.

An intriguing question that remains is why, in the temperature regimes investigated here, the array appears to act like a single aperture, or a single weak link. The amplitude of the oscillations in the phase slip regime, but close to the transition to weak coupling behavior, indicates that all the apertures are acting together[2]. One simple argument for phase coherence across the array is that phase gradients parallel to the wall containing the apertures correspond to lateral currents, which are not energetically favorable. It has been suggested that whereas thermal fluctuations can be strong in a single aperture, they may be suppressed in an array[12]. It is not at all clear in the strong coupling limit how the apertures interact and give rise to the synchronous generation of phase slips. We are working on extending these measurements to lower temperatures, and there is preliminary evidence that the array becomes less synchronous as T drops.

We find that the measured $I_s(\varphi)$ is well described by an empirical model consisting of a purely linear kinetic inductance in series with an ideal (purely sinusoidal) weak link. For the latter, $I_s(\theta_1) = I_c \sin(\theta_1)$. For the linear inductance, $I_s(\theta_2) = \hbar \theta_2 / m_4 L_\ell$. Here $\theta_1$ is the phase across the ideal weak link, $\theta_2$ is the phase across the linear inductance $L_\ell$, and $\varphi = \theta_1 + \theta_2$. The model can be characterized in terms of $I_c$ and the ratio of two inductances, $\alpha = L_\ell / L_J$. Here $L_J$ is the kinetic inductance of the ideal weak link evaluated at $\theta_1 = 0$, $L_J = (\hbar / m_4)(dI_s / d\theta_1)^{-1}_{\theta_1=0} = \hbar / m_4 I_c$. The overall current-phase relation can be written parametrically: $I_s = I_c \sin(\theta_1)$, $\varphi = \theta_1 + \alpha \sin(\theta_1)$. An analogous model has been applied to superconducting Josephson junctions[13]. It has been found to be inapplicable to $^3$He weak links[14]. The model parameters can be determined from the measured $I_s(\varphi)$ in a very simple way: $I_c$ is the maximum of $I_s(\varphi)$, which occurs at $\varphi = \varphi_m$, and $\alpha = \varphi_m - \pi/2$ is the deviation of the peak position from $\pi/2$. In the limit $\alpha \to 0$, the linear inductance is negligible and $I_s(\varphi) = I_c \sin(\varphi)$. In the limit $\alpha \gg 1$, $I_s(\varphi)$ is linear except near $\varphi_m$. For $\alpha \geq 1$, there exists a critical phase $\varphi_c$ at which $dI_s/d\varphi = -\infty$. The model is multiple valued in this case, and a phase slip occurs when the system falls off the cliff at $\varphi_c$ onto an adjacent branch of $I_s(\varphi)$. When $\alpha \gg 1$, $\varphi_c \cong \varphi_m$ and the size of the phase slip is $\Delta I_s = 2\pi I_c / \alpha$. The transition between the discontinuous phase slip regime and the continuous weak coupling regime occurs when $\alpha = 1$, $\varphi_m = \pi/2 + 1$, $\varphi_c = \pi$, and $\Delta I_s = 0$.

The measured parameters $I_c$ and $\alpha$ (independent of any model) are plotted versus temperature in Fig. 4 a and b. The model prediction for $I_s(\varphi)$, using only the measured $I_c$ and $\alpha$ values, is plotted in Fig. 4c for four different temperatures, along with the actual measured $I_s(\varphi)$. The agreement is striking, and shows that, within the temperature range we have investigated, the entire $I_s(\varphi)$ can be accurately reproduced by this model at a given temperature from the measured $I_c$ and $\alpha$ alone. Although the model is empirical, it lends insight into how the evolution of $I_s(\varphi)$ can be viewed as the transition from a multiple-valued hysteretic function to one that is single-valued.

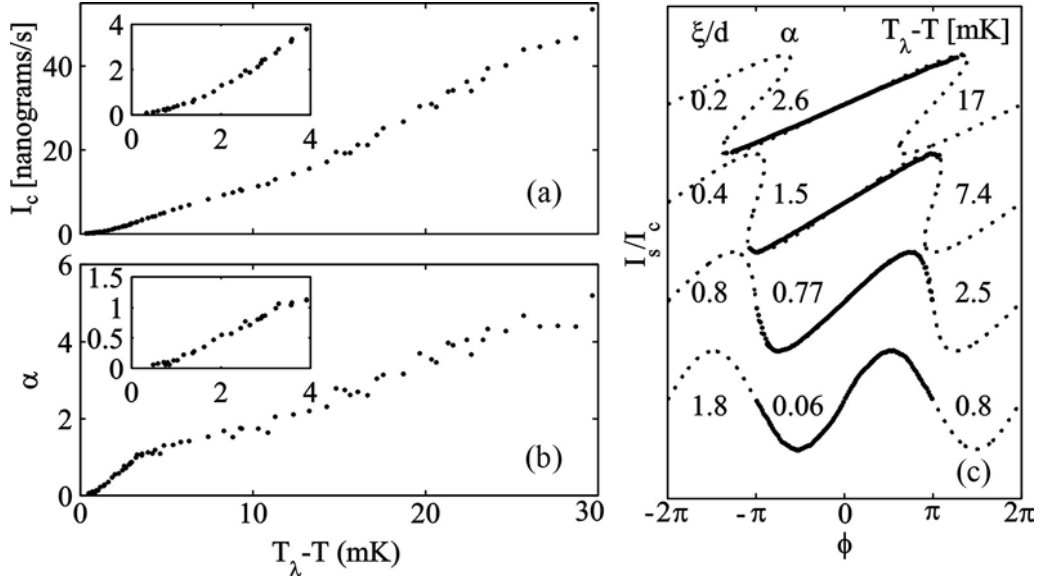

**Figure 4**. Evolution of superfluid $^4$He weak link parameters with temperature and an empirical model which can generate $I_s(\varphi)$ from these parameters. (a) Measured maximum current. (b) Measured deviation of $I_s(\varphi)$ peak position from $\pi/2$. The insets show an expanded view of the data within 4 mK of $T_\lambda$. (c) Normalized $I_s(\varphi)$, measured (solid points), and generated using model (dotted lines).

The experiment described here reveals the evolution of the function $I_s(\varphi)$ characterizing the union of two superfluid $^4$He reservoirs. This evolution shows a transition between two important and distinct quantum phenomena: phase slips, associated with the generation of singly quantized vortices, and the Josephson effect, associated with the weak coupling of two quantum systems through a potential barrier. We find that a simple two parameter model accurately describes the entire temperature regime under study. The $\sin(\varphi)$ behavior revealed at the higher temperatures will lead to the development of a superfluid $^4$He interferometer, an analog of the superconducting dc-SQUID. Such a device, operating near 2K, a regime accessible by mechanical cryo-coolers, will lead to practical devices useful in inertial navigation, geodesy and basic physics.

Correspondence and requests for materials should be addressed to R.E.P. (rpackard@berkeley.edu).

**Acknowledgements**

Dr. T. Haard made important contributions to the construction and design of the apparatus. M. Abreau also helped with the construction of the apparatus. The aperture arrays were fabricated by A. Loshak. We thank D. H. Lee and H. Fu for helpful discussions. This work was supported in part by the NSF (grant DMR 0244882) and NASA.